\documentclass[sigconf]{acmart}

\AtBeginDocument{%
  }

\newcommand{\add}[1]{#1}  
\newcommand{\del}[1]{}   

\copyrightyear{2025}
\acmYear{2025}
\setcopyright{cc}
\setcctype{by}
\acmConference[CHI '25]{CHI Conference on Human Factors in Computing
Systems}{April 26-May 1, 2025}{Yokohama, Japan}
\acmBooktitle{CHI Conference on Human Factors in Computing Systems (CHI
'25), April 26-May 1, 2025, Yokohama,
Japan}\acmDOI{10.1145/3706598.3713668}
\acmISBN{979-8-4007-1394-1/25/04}
\begin{document}

\title{How Scientists Use Large Language Models to Program}

\author{Gabrielle O'Brien}
\affiliation{%
  \institution{University of Michigan School of Information}
  \city{Ann Arbor, Michigan}
  \country{United States}}
\email{elleobri@umich.edu}

\begin{abstract}
Scientists across disciplines write code for critical activities like data collection and generation, statistical modeling, and visualization. As large language models that can generate code have become widely available, scientists may increasingly use these models during research software development. We investigate the characteristics of scientists who are early-adopters of code generating models and conduct interviews with scientists at a public, research-focused university. Through interviews and reviews of user interaction logs, we see that scientists often use code generating models as an information retrieval tool for navigating unfamiliar programming languages and libraries. We present findings about their verification strategies and discuss potential vulnerabilities that may emerge from code generation practices unknowingly influencing the parameters of scientific analyses.  
\end{abstract}




\maketitle

\section{Introduction}

"Software is as important to modern scientific research as telescopes and test tubes," writes \citet{Wilson2012BestComputing}. By some estimates, over 90\% of scientific researchers make or use software in fields ranging from  physics to biology to psychology \cite{Nangia2017TrackResearch, Hettrick2014ItsResearchers}. Programming powers critical activities like data collection, processing, analysis, and visualization \cite{NationalAcademiesofSciencesandPolicyandGlobalAffairsandBoardonResearchDataandInformationandDivisiononEngineeringandPhysicalSciencesandCommitteeonAppliedandTheoreticalStatisticsandBoardonMathematicalSciences2019ReproducibilityScience,Hey2009TheDiscovery,Baxter2006ScientificWww, Prabhu2011AScience}. The integrity of scientific results depends on code written and reused by researchers, and errors in even a single line can invalidate seemingly sound conclusions (see retracted works such as \cite{Karraker2015AuthorsRetraction, Mandhane2024Notice202317711:1226-1228., 2016Retraction:2016}).

Despite the central role of programming in modern research, scientists report insufficient training and incentives to master foundational development skills \cite{Hettrick2014ItsResearchers, Carver2022AStates, Wilson2012BestComputing, Prabhu2011AScience, Howison2010ScientificCollaboration, Howison2011ScientificCollaboration, Howison2015UnderstandingMeasures, OBrien2024InResearch}. By some estimates, around half of scientists who program have never received formal training \cite{Hettrick2014ItsResearchers, Carver2022AStates} \add{(though there are growing movements to offer targeted training for scientists \cite{Wilson2016SoftwareLearned, Carver2022AStates} and professionalize research software engineering \cite{Baxter2012TheEngineer,Sims2021ResearchIdentity})}. Even among experienced scientific programmers, quality control practices like software testing (i.e., unit and regression testing) that are common in professional development contexts are used unevenly, if at all \cite{Kanewala2014TestingReview, Hannay2009HowSoftware, Heaton2015ClaimsReview}. Because the integrity of scientific results often depends on code developed by programmers with little formal training, tools that support scientists who code are of tremendous importance.

One of the most notable breakthroughs in developer tools in recent years is code generation via large language models (LLMs), statistical models that are sensitive to long-ranging dependencies in text sequences like code and natural language. Here, we will use the term \textit{"Code LLM"} to refer to the use of any LLM for programming support. Code LLMs have shown promise solving common programming problems \cite{Chen2021EvaluatingCode, Li2022DebuggingProgrammers, Yetistiren2022AssessingGeneration}, and studies of early adopters suggest that less experienced programmers may especially benefit \cite{Ziegler2022ProductivityCompletion, Cui2024TheDevelopers}. Generally, programmers interact with Code LLMs via natural language chat (i.e., OpenAI's browser-based ChatGPT \cite{OpenAI2022ChatGPT:Dialogue}) or within their interactive development environment (IDE) via in-line code suggestions (i.e., GitHub Copilot \cite{Friedman2022IntroducingProgrammer} and TabNine \cite{Weiss2022AnnouncingModels}). 

Early studies have established a trade-off for Code LLM users: generated code must be checked for correctness, shifting the work of programming from drafting code to verifying it \cite{Chasins2023AnCode, Gu2023UnderstandingAnalysis, Barke2023GroundedModels}. Debugging generated code may be especially difficult, as it must be carefully inspected and understood \cite{Barke2023GroundedModels, Imai2022IsStudy, Ziegler2022ProductivityCompletion,Liang2024AChallenges,Nam2024UsingUnderstanding}---a non-trivial task for inexperienced programmers \cite{Nguyen2024HowOther, Dakhel2023GitHubLiability, Dak}. 

Because scientific conclusions can sensitively depend on how scientific programs are written, and given the uneven landscape of training and development practices in science research, the reliability and trustworthiness of scientific conclusions may be vulnerable to unintended programming choices introduced by Code LLMs. In this study, we make an initial inquiry into how Code LLMs are being used by scientists who program so we can better understand the vulnerabilities that may be introduced, as well as directions for tool design that can support code validation in scientific settings. 

\subsection{Approach}
This research makes three main contributions. First, we seek to understand \textit{who} is using Code LLMs among science researchers at a major university and what kind of interface they use (browser-based chat, like ChatGPT, or IDE-integrated autocomplete like GitHub Copilot). For brevity, we will call these two interfaces \textit{"Chat"} and \textit{"Copilot"} respectively throughout the manuscript. We are particularly interested in scientists' interface preference because each may have distinct ramifications for use cases and verification practices: in-IDE code assistants optimized for autocompletion generate shorter responses on average than models trained for natural dialogue \cite{Weber2024SignificantModels}. As such, we might expect Copilot-offered code suggestions to be easier to inspect and prone to fewer errors than long Chat generations involving both code and plain-language explanations (which are both known to be hallucination-prone \cite{Kabir2024IsQuestions}). Furthermore, popular Chat tools often require the user to context-switch between their development environment and browser, which may discourage using them in "acceleration-mode" (a term coined by \citet{Barke2023GroundedModels} to describe accelerating the implementation of already-planned code). Instead, users may default to using Chat in contexts where they are unsure how to implement something in code---a context in which verification may be considerably more challenging. Through a survey of $n=199$ researchers who program with Code LLMs, we find that the majority of respondents are using Chat interfaces instead of Copilots. These users are especially common in the life sciences rather than traditionally computational domains, like engineering and computer science.

Second, we ask, "Why do researchers use Code LLMs?" From interviews with $n=14$ users covering a variety of research areas and programming objectives, we find overwhelmingly that researchers program in multiple languages and libraries which they may have only surface-level knowledge of. This is an adaptation to the patchwork nature of software choices across and within labs. Code LLMs serve in place of documentation, enabling researchers to discover functions and methods that are available within unfamiliar programming contexts. 

Finally, we ask how scientists verify generated code. By combining interview responses with user logs from on-the-job interactions with Code LLMs, we find that the most commonly reported strategy is running generated code and visually inspecting its output. Some participants described circular strategies, like deferring to the perceived authority of LLM-generated explanations without referencing an external source like documentation. Participants were typically confident they could catch errors, but user logs revealed several instances where unintended behaviors were introduced by Code LLMs that could have scientific ramifications. 

Based on our findings, we see evidence that some researcher's code verification strategies may be sub-optimal for working with Code LLMs in a scientific setting, particularly in cases where visual inspection of computational results is difficult. This may be exacerbated by use of browser-based chat interfaces, which can produce and modify long blocks of code that are hard to fully understand, and by user's erroneous beliefs about how these models work.

\section{Background and Related work}

\subsection{How programmers use Code LLMs}
There is a growing literature on how software developers interact with Code LLMs, particularly focusing on in-IDE tools customized for programming like GitHub Copilot. \add{A frequent finding is that programmers use the autocomplete functionality of tools like Copilot to speed up the process of drafting code, particularly when writing formulaic or boilerplate structures \cite{Liang2024AChallenges, Barke2023GroundedModels,Sarkar2022WhatIntelligence,Vaithilingam2022ExpectationModels, Mozannar2024ReadingProgramming}. Programmers also use Code LLMs to reduce the cognitive load of remembering software libraries' API methods and language-specific syntax, drawing comparisons between the role of Code LLMs and Stack Overflow \cite{Liang2024AChallenges, Sarkar2022WhatIntelligence,Nam2024UsingUnderstanding,Barke2023GroundedModels,Vaithilingam2022ExpectationModels,Kabir2024IsQuestions}. Additionally, programmers report using Code LLMs to "provide a good starting point to approach a programming task” \cite{Vaithilingam2022ExpectationModels} that the user struggles to decompose into computational steps \cite{Sarkar2022WhatIntelligence, Liang2024AChallenges,Barke2023GroundedModels}. }

One close study of experienced programmers working with Code LLMs suggests these varied objectives map to at least two "modes" of interaction \cite{Barke2023GroundedModels}:  \textit{acceleration}, in which programmers know what they intend to write and Copilot helps them write it faster, and \textit{exploration}, in which they consider multiple possibilities for how to achieve a programming objective using Copilot to facilitate exploration. In acceleration mode, users prefer short, granular autocomplete suggestions that anticipate the design choices they have already made. In exploration mode, users deliberately prompt Code LLMs to suggest unfamiliar coding patterns, libraries, and APIs. 

\add{Considering the growing adoption of Code LLMs, it can be tempting to view them as a new form of programming that allows users to code via a higher level of abstraction than before--via natural language. If this is correct, Code LLMs should be the ultimate support for non-expert programmers like scientists. However, \citet{Sarkar2022WhatIntelligence} cautions that "a prompt can span the gamut from describing an entire application in a few sentences, to painstakingly describing an algorithm in step-by-step pseudocode. Thus it would be a mistake to view programming with AI assistance as another rung on the abstraction ladder. Rather, it can be viewed as a device that can teleport the programmer to arbitrary rungs of the ladder as desired.” }

\add{Programming with Code LLMs still requires users to understand and interact with code at multiple levels of abstraction due to Code LLMs' inherent unreliability: even well-crafted prompts can validly correspond to a variety of code solutions, which may be an ineradicable aspect of specifying program requirements via natural language \cite{Gu2023UnderstandingAnalysis, Sarkar2022WhatIntelligence,Ragavan2022GridBook:Grid,Liu2023WhatModels,Xu2022In-IDEChallenges}. Furthermore, Code LLMs are by design statistical models of sequences rather than algebraic engines \cite{Bender2021OnBig,Mirzadeh2024GSM-Symbolic:Models}. Their reliability in coding tasks depends on conditional factors like the quality and coverage of training data and the length and complexity of prompts \cite{Wang2024WhereCode,Chen2021EvaluatingCode,Yetistiren2022AssessingGeneration, Dakhel2023GitHubLiability,Lai2023DS-1000:Generation}. For the foreseeable future, programming with the assistance of a Code LLM requires the user to check that its outputs are acceptable---a task that requires both skill and effort \cite{Chasins2023AnCode,Mozannar2024ReadingProgramming, Barke2023GroundedModels}.}
 
Indeed, the ability to verify generated code appears to interact with the user's prior programming experience: a Copilot user study with less-experienced participants found that half reported difficulty understanding generated code, which made repairing errors and checking code correctness challenging \cite{Vaithilingam2022ExpectationModels}. Out of 24 participants, the experimenters also observed 8 cases where the participant accepted generated code without any effort to validate it. Despite this, a large majority of participants (19 of 24) indicated a preference to use Copilot in their regular programming practice. Similarly, \citet{Dakhel2023GitHubLiability} analyzed Copilot’s solutions to a battery of programming tasks and concluded that “…Copilot can become a liability if it is used by novice developers who may fail to filter its buggy or non-optimal solutions due to a lack of expertise”.  

Surveys of early adopters of Code LLMs have similarly confirmed that understanding, debugging and validating generated code is a major hurdle, even for professional developers \cite{Russo2023NavigatingEngineering, Liang2024AChallenges}. In response to this, a research team proposed and scoped an in-IDE tool for generating code explanations \cite{Nam2024UsingUnderstanding}. Unexpectedly, they found that more professional developers benefited more than student developers. In line with \citet{Vaithilingam2022ExpectationModels}, the researchers noted several instances where participants chose not to try to understand generated \add{code} at all---a phenomenon they called "outsourcing comprehension to the LLM". 

This literature establishes that on a variety of programming tasks, validating generated code is challenging even for software professionals. It is yet to be established how well these findings about programming generally apply to scientific contexts, though. To use the language of researcher Diane Kelly, the space between scientific programmers and software engineers is a "chasm" \cite{Kelly2007AComputing}: in addition to an array of cultural distinctions owing to radically different development incentives \cite{Hannay2009HowSoftware, Storer2017BridgingProgramming}, scientists may not see value in standard engineering practices \cite{Carver2013Self-perceptionsEngineers}, may not adopt explicit verification strategies like code testing, \cite{Kanewala2014TestingReview, Heaton2015ClaimsReview}, may never receive formal training in software development \cite{Carver2022AStates,Hettrick2014ItsResearchers}, and may not have clear upfront requirements for software when they begin developing \cite{Segal2009SomeScientists, Smith2019DebunkingSoftware}. Considering these distinctions, scientists will need to be studied as a distinct user group to understand how they interact with Code LLMs.

\subsection{How data analysts use Code LLMs}
Professional data analysts and data scientists have similarities to science researchers: they have all been classified as "end-users", programmers who seek not to develop code as a primary product but for a domain-specific goal \cite{Segal2007SomeDevelopers,Rothermel2011TheEngineering} (though this view has been challenged considering the breadth of development goals modern scientists may have \cite{Kelly2015ScientificSoftware}). Professional data scientists may use programming languages that are common in the sciences, such as Python and R \cite{Albright2018ACS1}, though like science researchers they often lack formal training \cite{Kross2019PractitionersAcademia, Kandel2012EnterpriseStudy}. Considering these similarities, the literature on how data professionals interact with code-generating tools is relevant. 

\add{While general findings from programmer interactions with Code LLMs likely still apply, distinct research considerations should be made for this population: software libraries used by data professionals, and the types of programming problems they encounter, may not closely match the training data of popular Code LLMs \cite{Lai2023DS-1000:Generation}. These users often work in specialized development environments, like Jupyter Notebooks, that interleave code, documentation, and visualizations; in-notebook interfaces for Code LLMs is therefore an active area of research \cite{Chandel2022TrainingAssistant,Yin2022NaturalNotebooks, Gu2023UnderstandingAnalysis,McNutt2023OnNotebooks}. Specialized Code LLM interfaces have also been explored for spreadsheet computation \cite{Ragavan2022GridBook:Grid} and data visualization \cite{Dibia2023LIDA:Models}.}

\add{Code LLMs also hold unique promises for data work. For example, they may lower barriers for users to implement a variety of computational techniques \cite{Gu2023UnderstandingAnalysis}, circumventing an established bottleneck for many analysts \cite{Liu2020UnderstandingPractices}. Automated approaches that help analysts rapidly explore a broad landscape of modeling decisions could lead to more robust results, with less sensitive dependence on the idiosyncrasies of the analyst's strategy  \cite{Archambault2024SomeAnalysis}. At the extreme, though, \citet{Archambault2024SomeAnalysis} cautions that over-automation could defeat the goal of analysis in the first place: "Without any visibility into how [an interpretation of data] was produced, the human has no opportunity to apply knowledge that is not contained in the training data to debug operations chosen by the machine. They may not feel confident in their results, and their lack of insight into how they were reached may prevent them from applying the knowledge that is output, leading to a question of whether it is knowledge at all."}

\add{Considering the careful relationship between data analyses and knowledge production, it is perhaps unsurprising that user studies indicate complicated attitudes towards Code LLMs and a diversity of strategies for understanding their outputs.} One interview-design study of professional data scientists working in Jupyter notebooks found a variety of attitudes \add{regarding} Code LLM assistance, ranging from distrust of fully generated code to appreciation of its value as a teaching tool \cite{McNutt2023OnNotebooks}. Participants reported verification strategies like consulting external documentation or using  Code LLMs to generate multiple candidate suggestions, which could signal the reliability of an answer (a pattern also observed by \citet{Barke2023GroundedModels}). 

Another recent study of how analysts verify AI-generated data analyses \cite{Gu2024HowAnalyses} observed users engaging in a combination of \textit{procedure-oriented} and \textit{data-oriented} verification strategies---asking "what does this generated code do?" and "do data objects being manipulated and produced by code make sense?" Analysts with more professional experience appeared to make more explicit efforts at verification, such as formulating expectations about data objects and checking them \{add{(though these attempts are still different from formal code testing)}. \add{Interestingly, what \citet{Gu2024HowAnalyses} call "eyeballing" and "quick checks"---high-level pattern-matching behavior to inspect properties like the shape or column names of data objects---was reported to be a successful strategy for validating Code LLM generations. In a similar study on how analysts program via a natural-language interface for spreadsheet computation, \citet{Ragavan2022GridBook:Grid} noted that users following such a strategy were prone to overconfidence in incorrect results. For example, they observed a user accept a summary statistic table on the basis it had the correct shape, despite having incorrect numerical values. These divergent interpretations indicates there may not be a widely-held consensus about what activities constitute reasonable validation for data-centric programming with Code LLMs.}

It is unclear exactly how much data science coding resembles scientific coding. There are likely  differences stemming from organizational factors, commercial versus academic incentives, and the goals of analyses---for example, data science often follows a predictive modeling paradigm than an inferential paradigm where falsifying hypotheses is the goal \cite{Breiman2001StatisticalCultures, Donoho201750Science}. Additionally, there are some kinds of scientific programming that do not closely correspond to analysis-driven programming, such as numerical simulations of physical systems that are common in atmospheric sciences, biophysics, or ecology \cite{Joppa2013TroublingUse,Basili2008UnderstandingPerspective,Johanson2018SoftwareScience, Prabhu2011AScience}. It remains to be determined which Code LLM user patterns can be generalized from data analysts to scientists.

\subsection{How novice programmers use Code LLMs}
While many scientists who program are not novices, they often lack formal training \cite{Hettrick2014ItsResearchers, Carver2022AStates, Nangia2017TrackResearch}. As such, the rapidly accumulating literature on beginner programmers' interactions with Code LLMs may be relevant. 

\add{There has been tremendous excitement about generative AI as a programming teaching tool \cite{Liu2024TeachingEducation,Becker2023ProgrammingGeneration}. Yet studies so far suggest substantial variation in how effectively students use, understand, and learn from Code LLMs \cite{Prather2023ItsProgrammers,Prather2024TheProgrammers,Mordechai2024NoviCode:Novices,Jost2024TheOutcomes,Feldman2024Non-ExpertFuture,Margulieux2024Self-RegulationProblems}. } In one controlled study, 120 students who had completed a single introductory computer science class were tasked with solving Python problems with a custom Chat interface \cite{Nguyen2024HowOther}. Many found it difficult to construct natural language prompts that correctly and completely specified their intent. The authors hypothesize that students' difficulty refining a prompt strategy could relate to misconceptions about how Code LLMs work---in particular, they tended to believe the model employed rule-based keyword-lookup to retrieve information from a data structure. \add{Other studies about how novice programmers prompt suggest they misunderstand the scope of information Code LLMs can access \cite{Lucchetti2024SubstanceLLMs} and struggle to prompt at appropriate levels of abstraction required for successful code synthesis \cite{Liu2023WhatModels,Mordechai2024NoviCode:Novices, Feldman2024Non-ExpertFuture}. Beginners may also struggle to decompose programming problems into smaller tasks before prompting \cite{Jayagopal2022ExploringProgrammers,Liu2020UnderstandingPractices}, which is known to improve LLM performance on a variety of benchmarks \cite{Wei2022Chain-of-ThoughtModels,Patel2022IsNeed,Honovich2022InstructionDescriptions} }. 

\add{There is some evidence for a "widening gap" between students who can use Code LLMs in ways that productively scaffold their learning, and students who cannot \cite{Margulieux2024Self-RegulationProblems, Prather2024TheProgrammers,Jost2024TheOutcomes}. Close observational studies of beginners solving introductory programming tasks by \citet{Prather2023ItsProgrammers,Prather2024TheProgrammers} indicates that students who struggle to program may overestimate the correctness of generated code as well as the productivity gains associated with using them.  Students who exhibited maladaptive uses of Code LLMs in one study were prone to overstating their mastery of foundational programming concepts, potentially delaying their own learning.}

\section{A survey of scientists who use Code LLMs}

First, we present a survey of scientists who use Code LLMs. The target population was researchers at a large, public university who a) program as part of their scientific research,  and b) have tried using any generative AI code tools to support their programming. 

\subsection{Methods}

This study, as well as the interviews described in Section \ref{interview-study}, were reviewed and approved by the Institutional Review Board at the University of Michigan. The survey was conducted in May 2024. Participants were recruited via a targeted university email campaign. Using a list of emails provided by Human Resources, we contacted 8,876 employees with job titles classified as scientific research-related, including graduate student research assistants, post-doctoral researchers, research staff, and faculty with research appointments. Additionally, we advertised the study through posts in several relevant internal Slack channels, such as a campus-wide research computing support group.

Researchers were invited to complete a Google Forms survey about their research programming practices and experience with generative AI code tools in this context. The survey included questions about their job title, department, the primary programming languages they use in their scientific work, years of programming and research experience, which generative AI code tools they have tried, and how often they use them. Additionally, the survey included open-ended questions about the role of programming in their scientific work and how they use generative AI code tools. \add{Participants were not compensated for survey responses. However, participants were informed the survey would be used to recruit candidates for interviews, which would be compensated (Section \ref{interview-study})}. The complete survey is provided as supplemental material to this article.

\subsubsection{Preparing survey responses for analysis}
We received 230 unique responses. To ensure respondents were in the targeted population, we filtered out responses from any individuals who indicated their primary appointment was not at the university (15 responses), did not program at least once a month as part of their scientific work (6), and had never tried using a generative AI tool in their programming (8). We also removed responses that did not indicate what language they programmed in (1) or how often they used a Code LLM (1), as the absence of these responses made it difficult to confirm the respondent was in the targeted study population. This left a total of 199 responses. 

Before statistical reporting and analysis, a number of data transformation steps were performed: because the distribution of respondent's years of experience (for both science research and programming) had a long right tail produced by a small number of researchers with decades-long careers, these measures were log-transformed. This step reduces the propensity for any correlations with these variables to be driven by a handful of extreme values. 

To analyze programming language frequencies and co\-occurrences, we created binary variables for the most commonly mentioned languages (i.e., \verb |uses_python| which takes the value 1 if a participant reports using Python as part of their scientific work and 0 otherwise). As a threshold, we performed this step for any language reported by more than 3 respondents (Python, R, MATLAB, C++, Bash, Stata, SAS, JavaScript, FORTRAN, C, Julia, Java, Mathematica and SQL).

Finally, we used the National Science Foundation's codes for research classifications to map respondent's home departments to a research classification (for example, "Life Sciences", "Physical Sciences", "Engineering", "Social Sciences"). While the correspondence between a researcher's home department and their research agenda is imperfect, this standardized classification enables high-level reporting of the research areas represented in our survey responses.

\subsection{Results}

To contextualize our findings about programming practices and Code LLMs, we begin with descriptive statistics about the backgrounds of respondents. Our 199 respondents represent a range of academic disciplines, with the most common being Life Sciences (94) followed by Engineering (39) and Computer \& Information Services (27). Other research classifications included Geosciences (9), Psychology (9), Physical Sciences (8), Social Sciences (7), Law (2), Education (1), Mathematics (1) and Social Work (1). Participants tended to represent relatively early-career scientists in training positions, with 89 graduate students and 45 post-doctoral researchers. There were also 47 research staff and 18 faculty respondents. About 60\% of respondents reported their gender as male (122) \del{respondents described their gender as male}, with the remainder reporting their gender as female (72) or non-binary (4). 

\subsubsection{Programming experience}
While a range of programming backgrounds and experience levels were captured in the responses, the median respondent had 6 years of programming (SD = 6.4 years). Notably, years of programming experience was correlated with years of science research ($r = 0.45$, $p < 0.001$), suggesting that these activities are coupled for many respondents. 

By design, only scientists who reported programming at least daily (111), weekly (69), or monthly (19) as part of their research were included in our study. The most common programming languages scientists listed were Python (133), R (93), MATLAB (36), C++ (21), Bash (15), Stata (12), SAS (11) and JavaScript (11). A slight majority of scientists indicated using more than one language as a primary part of their research : 53\% of respondents (106) reported using more than one, and the median survey respondent listed 2 primary languages. 

\subsubsection{Code LLM usage}
All respondents reported having tried using a Code LLM for the purposes of supporting their scientific programming (as this was an inclusion criterion), but their self-reported usage patterns and tools of choice varied considerably. When asked to select the statement that best describes their usage, 91 respondents chose "I use them sometimes when programming", 75 chose "I use them most times I am programming", and 33 chose "I have experimented, but don't regularly use them." 

Participants were also invited to select any tools they had explored from a checklist, with the option to write in additional choices. The most common response was ChatGPT (180), followed by a university-provided version of ChatGPT (115), GitHub Copilot (53), Anthropic's Claude (11), HuggingFace's HuggingChat (6), TabNine (4), Google's Gemini (2) and Meta's Llama (2). Broadly, these tools use two styles of interfaces: a natural-language chat accessed via browser,  (e.g., ChatGPT, Claude, HuggingChat) or an auto-complete plugin for interactive developer environment, or IDE (e.g., GitHub Copilot and TabNine). Using the terminology previously introduced, we classify the browser-based tools as Chat interfaces and the in-IDE tools as Copilots. Only about 29\% of respondents (57) reported experience with Copilots. This is noteworthy considering that at the time of the survey, all participants would have had free GitHub Copilot licenses thanks to their university affiliation, and Copilot was available as a plugin for many of the IDEs commonly associated with the most used programming languages in our sample (such as Visual Studio Code, PyCharm, and RStudio). It is also curious that the overwhelming majority of responses indicated experience with only closed-source models, with the exceptions being HuggingChat and Llama. 

\subsubsection{Predictors of Code LLM usage frequency}
As an exploratory analysis, we looked for correlations between survey measures with particular attention to predictors associated with self-reported LLM usage frequency. Because our survey sample consists of scientists who self-report being regular programmers with experience using at least one Code LLM, this analysis only considers factors that may distinguish casual users ("I have experimented, but don't regularly use them") from consistent users ("I use them most times I am programming"). 

\begin{figure*}
    \centering
    \includegraphics[width=0.5\linewidth]{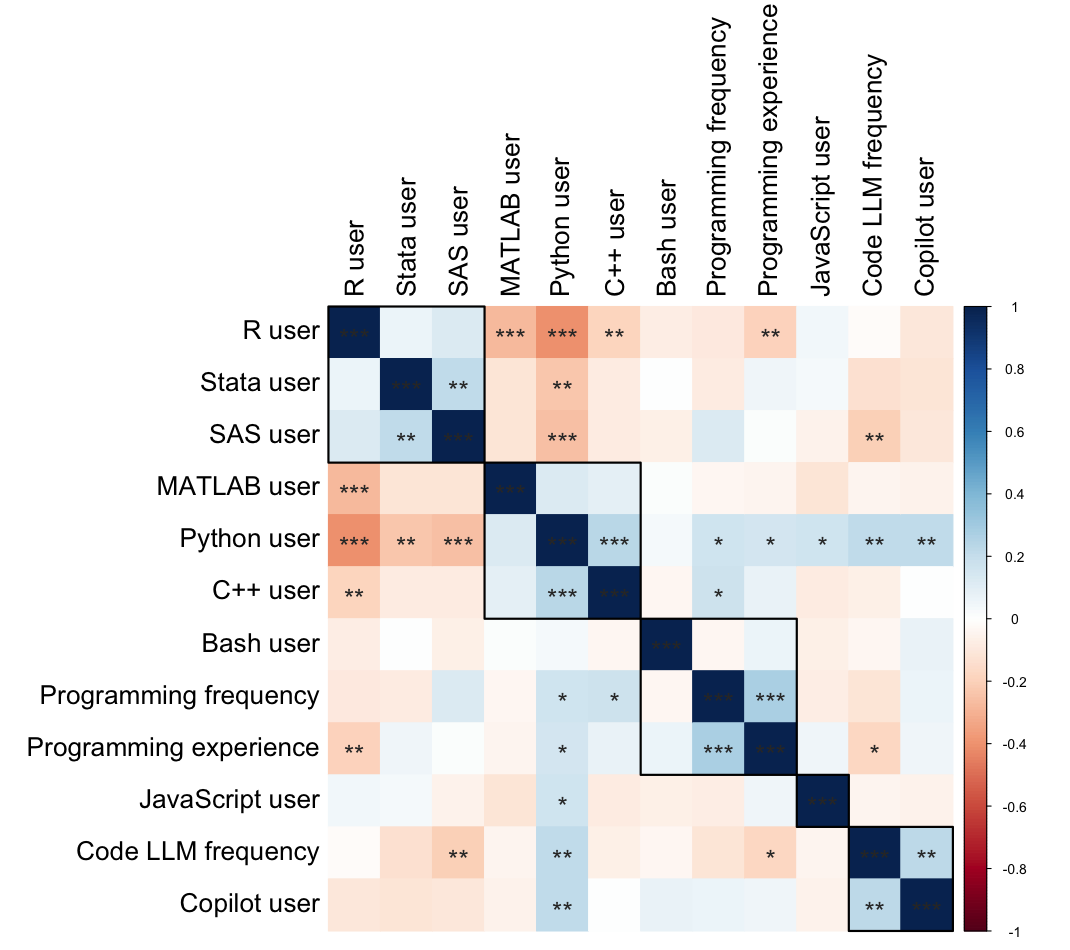}
    \caption{Correlation matrix for programming-related survey variables. Stars indicate statistical significance at the 0.05 (*), 0.01 (**), and 0.001 (***) levels. All $p$-values are shown uncorrected. Black rectangles indicate clusters of predictors determined by hierarchical clustering with 5 clusters.}
    \label{fig:corr-mat}
\end{figure*}

Figure \ref{fig:corr-mat} shows correlations between the survey variables related to programming habits (note that only languages with more than 10 users were included for statistical power). In order to surface patterns, an unsupervised learning technique, hierarchical clustering, was applied to sort predictors into discrete groups based on the strength of correlations between predictors. 5 clusters were determined to be the best fit on the basis of two measures of fit for cluster assignments, the Silhouette score and within-cluster-sum of squared errors \cite{Rousseeuw1987Silhouettes:Analysis}. The first cluster relates to using the R, Stata, or SAS programming languages; positive correlations between these predictors indicated that these languages are used by many of the same survey respondents. Similarly, a cluster relating to being a MATLAB, Python, or C++ user suggests overlap in the use of those languages by a distinct group of respondents. A third cluster consists of self-reported programming frequency at work, years of programming experience, and being a programmer who writes Bash scripts. A variable indicating the use of JavaScript is sorted into its own cluster (though it appears to have a positive but small correlation with Python use). Finally, a cluster contains the variables representing self-reported Code LLM usage frequency and using a Copilot interface. It is perhaps unsurprising that these variables have a moderate positive relationship, considering that a browser-based Chat interface has to be deliberately accessed outside the programming environment, whereas an in-IDE Copilot is always present. 

\subsubsection{Relationships with research field}
We also considered whether there were  differences in the frequency and mode of Code LLM use by field (Life Sciences, Computer \& Information Services, Engineering, or Other). 

Computer \& Information had the highest rate of Code LLM users who "use them most times" when programming at 59.3\% of respondents, compared to 35.9\% from Engineering,  27.7\% from Life Sciences, and to 48.7\% from all other fields. Engineering, as well as Computer \& Information, had the highest rates of Copilot use (41.0\% and 40.7\%) compared to only 20.2\% of respondents from Life Science and 23.1\% from Other fields. 

\subsubsection{Interim summary of findings}
The modal respondent to our survey is a Life Sciences researcher who accesses Code LLMs primarily via general-purpose, browser-based Chat interfaces, rather than via IDE-integrated tools like GitHub Copilot. Respondents are frequently bilingual programmers who use Code LLMs some of the time. Exploratory analysis suggests that reported usage frequency may vary by programming language, with the highest rates associated with Python use. Our results also suggest there is a subset of frequent Code LLM users in Computer \& Information and Engineering, who are especially likely to be Copilot users. 

\section{Interviews with users} \label{interview-study}
To better understand the uses of Code LLMs in scientific programming and researchers' motivations for adopting them, we recruited scientists who responded to the survey for semi-structured interviews about their experiences. We also invited them to share artifacts---exemplar interaction logs with Code LLMs from their own research practice---to align interviewee's self-descriptions of usage with their actual prompts and usage patterns.

\subsection{Methods}
\subsubsection{Recruitment}
We selected interview candidates from the survey respondents to maximize the diversity of potential Code LLM use-cases. In addition to considering the survey variables reported previously, we also examined responses to two open-ended questions: "What is the role of programming in your scientific work?" and "What do you use Code LLMs for in your research programming?" As such, we invited participants on a rolling basis to obtain a range of programming experience and research roles, as well as coverage of major programming languages indicated by the survey (details presented in Table \ref{tab:interview-participants}). Additionally, we recruited participants with a variety of programming objectives, including developing software that controls data-collection apparatuses (like behavioral experiments administered over a browser or a set of cameras), statistical modeling (including classical statistical inference as well as machine learning applications), designing and exploring parameters of numerical simulations (such as mathematical models representing physical phenomenon), and developing scientific software for communal reuse (either within the research group or with the broader research community). The frequency of these various programming objectives are reported in Table \ref{tab:programming-objectives}. Participants were invited to participate via email on a rolling basis until saturation was reached---that is, the interviews began to confirm themes and findings from previous interviews, rather than generate new ones \cite{Saldana2016The14}. In total, 23 invitations were sent resulting in 14 completed interviews. 

\subsubsection{Interview procedures} 
Semi-structured interviews were based on an interview guide, a set of questions that provide a scaffold for conversations, with follow-up about noteworthy responses on the judgment of the interviewer. At a high-level, every conversation was intended to address two overarching questions: 1) "Why do you use Code LLMs in your scientific research?", and 2) "How do you verify that generated code is acceptable?" Based on themes observed during the first two interviews, we added questions about participants' mental models of Code LLMs--specifically, how they understand them to work and what use cases they would consider acceptable. The interview guide is provided as supplemental material. Interviews typically lasted between 45-60 minutes and were conducted via Zoom. Recorded videos were transcribed for analysis using Rev, a commercial service. Participants were compensated \$25.

\subsubsection{Artifact collection}
While scheduling interviews, we asked researchers to provide sample logs from Code LLMs produced during their research-related programming. Some participants provided these before the interviews, whereas others followed up via email afterward to provide logs for specific use cases they had recounted during the interview. In all, 10 participants provided chat log files and 1 participant screenshared during the interview to demonstrate using GitHub Copilot in his IDE (P13). 

\subsubsection{Analysis} 

Interview transcripts were analyzed \add{by the same author} on a rolling basis with several coding cycles in Atlas.ti, a software for qualitative analysis. First, open coding was conducted, marking themes of key quotes ("documentation", "looking up functions", "reading line by line"). After 10 interviews, \add{the same author conducted axial coding} to consolidate similar open codes and create a taxonomy. Here an inductive approach was followed \cite{Corbin2012BasicsTheory} to categorize codes into themes like "use cases", "verification strategies", and "mental models". This scheme was \add{applied by the coder for the final 4 interviews}, with minor refinement as needed.

\begin{table*}[]
  \caption{Interview participant details.}
\begin{tabular}{rlllll}
\toprule
\textbf{ID} & \textbf{Field}        & \textbf{Role}  & \textbf{Primary Languages}      & \textbf{\shortstack{Code LLM\\interface}} & \textbf{\shortstack{Years exp.\\coding}} \\
\midrule
P1                              & Social sciences        & Research staff & Python, JavaScript      & Chat                  & 1                                                           \\
P2                              & Psychology        & PhD student    & Python, R, JavaScript   & Chat                  & 6                                                           \\
P3                              & Life sciences         & Research staff & STATA, R                & Chat                  & 3                                                           \\
P4                              & Life sciences       & PhD student    & Python                  & Copilot          & 8                                                           \\
P5                              & Life sciences         & PhD student    & SAS, R                  & Chat                  & 6                                                           \\
P6                              & Social sciences        & Research staff & R, Python               & Chat                  & 6                                                           \\
P7                              & Life sciences       & Research staff & R, Python               & Chat                  & 2                                                           \\
P8                              & Physical sciences             & PhD student    & C++, Python, MATLAB     & Chat                  & 8                                                           \\
P9                              & Computer \& Information       & PhD student    & Python                  & Chat                  & 5                                                           \\
P10                              & Life sciences        & PhD student    & Python, Java            & Chat                  & 8                                                           \\
P11                              & Life sciences        & PhD student    & R, STATA                & Chat                  & 5                                                           \\
P12                              & Geosciences & Faculty        & Python, Matlab, FORTRAN & Chat                  & 22                                                          \\
P13                              & Engineering           & Faculty        & Python, R, Mathematica  & Copilot          & 50  \\
P14                              & Life sciences           & Post-doctoral researcher        & Python, R  & Chat          & 2  \\
\bottomrule
\label{tab:interview-participants}
\end{tabular}
\end{table*}

\begin{table}
\caption{Scientific programming objectives among interview participants}
    \centering
    \begin{tabular}{ll}
    \toprule
    \textbf{Programming objective} & \textbf{Frequency} \\
    \midrule
         Data collection & 5 \\
         Statistical modeling & 13 \\
         Visualization & 13 \\
         Numerical simulations & 5 \\
         Software library development & 3 \\
    \end{tabular}
    \label{tab:programming-objectives}
\end{table}

\subsection{Results} 

\subsubsection{Why do these researchers use Code LLMs?}

Across interviews, a dominant theme emerged around Code LLMs as a tool for information seeking about the functions and methods available within programming libraries, example use cases for those functions, 
and the meanings of unfamiliar functions encountered in code written by other people (like former labmates). Many participants explicitly recounted how their tool of choice had supplanted traditional search engines like Google, searchable user forums like Stack Overflow, and official documentation associated with programming libraries. "Every question that I have, I put into ChatGPT," said P11. "Whether it's how do I extract a variable in a column to do X, or much more complicated tasks such as how do I adapt the format of a graph." 

But what motivates the need for information retrieval in the first place? Many interviewees attributed their information seeking to low fluency with one or more of the programming languages (or software libraries) they used in research code, which occurred for several reasons. 

In some cases (P6, P8, P11, P14), the researcher had to learn a new programming language upon joining a lab with a dominant language different from the one they had trained in. P11 had only ever programmed in R before accepting a research staff role, but had to adopt Stata to work with the principal investigator.  Similarly, P14 had some experience in Python, but recalled "when I joined [the lab], they do everything pretty explicitly in R. I just had to shift over to use their system.” For P6, the need to adapt to a new language was driven by conventions established by a previous lab member: “I 99\% use R, but I have inherited Python code from the previous research assistant, so I have to understand Python code.” One interviewee (P1) learned a combination of Python and JavaScript programming entirely on the job with no prior experience in any language.

For other interviewees (P2, P3, P9), accommodating collaborators drove adoption of new programming languages. As P3 explained, "I'm with some PhD students that use R, so I switched to R because it's code they know." 

Several interviewees mentioned intentionally using a combination of languages for different programming objectives, based on the libraries available in each. P5 preferred to process and analyze data in SAS but visualize in R. Similarly, P9 worked in Python except when doing causal inference, as they preferred the packages available in R. For P10, whose research involved a combination of developing open source tools with graphical user interfaces for his scientific field as well as analyzing data within the lab, a combination of Java and Python made sense: "I use Java for building the tools and Python for the data analysis and visualization."

The interviewees with the longest careers, P12 and P13, both reported changing their primary programming language in response to changing norms in their fields. For P12, FORTRAN was phased out in favor of MATLAB and eventually Python in their professional community. They spoke of lacking a certain intuition for Python conventions even several years after switching: "I have to admit, I still don't find a lot of it intuitive in terms of when do you use brackets, when do you use curly brackets, when do you use parentheses?" 

 P13 similarly described transitioning from Mathematica to Python for most objectives (except using R for statistical analysis), motivated by the open source movement. Day to day work for these senior researchers could involve a mixture of languages, as some components of a project might be written in a previously-preferred language. 

For all these reasons, it was common for interviewees to develop research code in languages and libraries where they could not readily recall the functions available, syntax requirements, and meaning of common error messages. This could occur from a lack of programming training in general, but among this sample \add{it} was more likely to stem from switching \textit{between} languages and libraries to contexts where the researcher lacked deep knowledge of a language's conventions and affordances. 

\subsubsection{Why did these researchers prefer Code LLMs to search?}

For interviewees who described their use of Code LLMs as replacing information retrieval via search engines, we asked what they preferred about Code LLMs. Several researchers expressed difficulty crafting an effective search query when facing unfamiliar syntax or error messages (P2, P5, P8, P10), and so found pasting their code or error message into a chat interface to be an easier starting point than writing a query. Others preferred that generated responses meant less time evaluating the relevance of multiple sources, such as the top few Google search results or Stack Overflow comments  (P3, P9, P10). "Google will return multiple results and they are sometimes not exactly what I want," said P9. P10 similarly said, "With Stack Exchange, if I Google something, I have to go through all these links that may or may not be relevant and it's often tangential, conventionally related." P10 also voiced a preference for receiving summaries of sources rather than the original source.

Research code can also involve niche elements that are uncommon outside of a scientific community, potentially limiting the number of relevant sources of information to be retrieved. P4 and P10 raised the possibility that there were no similar cases to their own on forums like Stack Overflow. For P2, working with a field-specific library for conducting experiments over the browser, there was limited documentation of errors and their causes to search. For P4, the documentation for Stata seemed incomplete: "They just don't provide relevant examples, that's a problem. And so that's where ChatGPT is useful because they will give you relevant examples, but Stata does not."

Some interviews touched on the role of official software documentation, which is designed as an authoritative reference. P12, a researcher with more than two decades of programming experience, described ChatGPT as "1,000 times easier [than documentation]. Just ridiculously easier."  For P1, who learned how to program on the job, documentation was initially unusable: "Until I had enough coding experience, I didn't even look at the documentation because I would open it, read it, and there were so many words that I didn't understand. So I just closed it... But these days I also don't read them because I realized I can ask ChatGPT to read them, and then apply the information to my code." P4, in contrast, continued to encourage students he supervised to use documentation as a first resource.  "I think the docs are better." He described his use of GitHub Copilot as "more of a time thing.”

\subsubsection{How do scientists reason about whether generated code is usable?}

\add{In response to questions like, "How do you determine if the output of your Code LLM is acceptable?" and "How you have you caught unacceptable outputs in the past?",} most interviewees expressed confidence they could screen generated code elements for correctness. They described a variety of strategies for verifying code.

The most common strategy reported was \textbf{running generated code and inspecting the outputs} (P1, P2, P5, P6, P7, P8, P9, P10, P11, P13, P14). This usually involved manually checking attributes of a data object produced by the code, such as visually reviewing values in a dataframe or properties like size and shape. In some cases, outputs could include artifacts such as browser console logs or video files, which have to be checked outside of the development environment. Outputs were sometimes checked with reference to an implicit or explicit standard. P9 described comparing outputs to her expectations: "I exactly know what the outcome I want it to be. For data cleaning step, this is just turning some structure of data into a different form. So it's very easy to tell." Similarly, P14 recalled a case that was "self-validating", in which she used ChatGPT to suggest code that could merge several cells in a dataframe into a string. In cases where researchers used Code LLMs to translate code between languages or refactor existing code, outputs could be explicitly compared between versions (P1, P2, P5, P6). 

Two researchers mentioned an explicit testing framework for checking code. "I test all my code because my code sucks....Whether it comes from Copilot or from my fingers, I assume it's wrong and I test it because otherwise I'm going to miss stuff," said P13. P8 had learned about testing from a senior research staff member, who had instilled "a culture in our lab around testing your code." P8 had a set of unit tests for one of their code projects in the lab, but not other projects. P13 described their process as less formal. "I don't necessarily do it in a structured way where those tests are going to become part of the code base because that hasn't been necessary in the kinds of things that I'm doing." They reported writing unit tests in specific cases where code would be used as a "black box" by another person, but "most of the time I'm just putting in different inputs that should work and verifying that the output is what I expect." 

A more general kind of output checking involved \textbf{observing runtime errors}.  P10 recounted they discovered that a generated code snippet passed a non-existent argument to a function when "I put it in Python and it just didn't run." P4 said, "Majority of time I'd just try it and then either it works or it doesn't." User logs from P1, P10 and P14 show back-and-forth interactions in Chat, in which the user continued to prompt the Code LLM with error messages produced by running generated code until the Chat suggested code that no longer produced errors. 

Several interviewees described \textbf{reading generated code line by line} as part of their verification process (P2, P3, P4, P13, P14). "I would just go line by line and make sure everything makes sense for what I'm doing," said P3. P6 said: "I basically read it myself. I don't work with NumPy and Pandas that much, but I have a basic understanding of their syntax, so I can read it and check that [ChatGPT]'s correctly implementing it." 

Some researchers described \textbf{looking for specific patterns in generated code}: “I guess you look at the control structure, hey, is there a forward loop or a while loop that looks like what we want to iterate over? Do the function calls look familiar? Do the arguments to the functions look familiar and sensible?" said P12. Similarly, P2 described their process with Chat, “I then take [the code], copy and paste it over to my IDE, swap out the variables I need, and I do just a quick just check like, "Okay, that if statement makes sense, this makes sense.”" P13 narrated how they thought through the correctness of a suggested code block during a screenshare with the interviewer, pointing out features like variable names, the names of of attributes attached to data objects, and the sequence of operations. 

Finally, several participants reported \textbf{referring to the LLM generated explanations} as a source of verification (P1, P7, P10, P11). When asked how he decided to use a function suggested by ChatGPT, P10 described being convinced because "it gave a good paragraph about it". In a similar context where P11 encountered an unfamiliar function in a generated code block,  "I asked it what is the function doing? And then it explains to me, the function is part of this package, and it solves linear programming problems. Then it  tells me how it works. And I was like, "Oh, okay, that makes sense."" This researcher also described "an iterative process where I'm trying to think with it whether the solution that it proposes makes sense to me. So I'm trying to make it explain it to me so that I understand what it's doing and when it makes sense to me and I see that it's working on a technical level and a conceptual level, then I'm convinced." P7 described using LLM-generated explanations to create a kind of "diff" between generated code versions: "If I don't want to look through [the code] particularly, I'll ask it, "What did you change in this version?" I'll have it tell me.” 

Beyond specific strategies, some interviewees mentioned the importance of developing an intuitive sense for when generated code is usable. P2 recalled trying to teach new students in the lab "to develop a sense of intuition of when it's wrong, when it's right. And that's very easy to say just in words, but it's very hard to communicate the exact when to do this and when not." P12 said of this intuition, "It's just a feeling of recognition, which is a fuzzy answer, but there is some kind of feeling of, "Yeah, that looks familiar, that looks pretty much like what I would do." P13, on the other hand, cited a lack of intuition about Code LLMs performance across development contexts as a concern: "You know if you're working with somebody, you develop a level of understanding of their competence. And when you worked with somebody for a long time, you know they can do that, they can't do this....And I don't have that yet for Copilot....the reason I don't have it is because it's really variable."

\subsubsection{Do these verification strategies work?} 
User logs indicated several instances where the user appeared to miss errors or unusual behaviors introduced by Code LLMs: 

\begin{itemize}
\item P2 prompted ChatGPT to generate a program corresponding to the signal processing steps listed in the methods of a scientific paper. Specifically, the program needed to rectify a time series, apply smoothing with a certain window size, and then deduplicate events within a certain window.  The model produced a $\sim$40-line working code example, but with the wrong smoothing window size (twice the desired length) and no event deduplication (while a code block was present with the comment "merge close bursts", its logic would result in no deduplication occurring). The researcher accepted this code and did not notice anything unusual about it until asked by the interviewer. \del{Interestingly, there was a full code base published alongside the paper that P2 was attempting to reproduce.}

\item P1 prompted ChatGPT first with a >1000-line script that executed an online, browser-based data collection protocol, then with two blocks of code that were intended to conduct an attention check for remote participants. All the code had been written by a previous employee who was no longer with the lab, and P1 described being unable to read the code when starting in the lab due to inexperience. The researcher prompted ChatGPT to choose which of the two attention check functions she "should use". ChatGPT responded that one was better than the other on the basis of code style and better error handling, and the indicated version appears in the script in follow-up \del{conversations} \add{prompts (i.e., the user continued to prompt ChatGPT with the full program several times in the same session}).  However, the two code blocks actually did different things---one instructed the user to raise the volume on their computer and try again. The other instructed them to try again (with no mention of volume) and allowed up to 2 more attempts before the program would advance. 

\item P14 prompted ChatGPT with $\sim$250 lines of code that controlled data collection from a camera and prompted it to add a new feature. ChatGPT returned a modified version of the script. The researcher realized the next day that no data had been collected from the camera during the night since adding the feature, as ChatGPT had quietly removed required logic that turned the cameras on. P14 recounted this experience during the interview and stated that afterwards, they changed their practice to avoid prompting ChatGPT with all their code at once. Curiously, other user interactions from around the same time indicate the researcher had encountered similar behavior before, where ChatGPT unexpectedly removed lines of code in addition to adding modifications. The choice to accept modified code without directly comparing it to the original may reflect expectations about effort: perhaps the work of understanding the modification seemed less than the work of troubleshooting issues with the new code. 

\end{itemize}

While our study does not allow us to estimate the rate of errors being missed by researchers in generated code \add{or code explanations}, the prevalence of these issues in our 14 interviews (with access to only a limited subset of user logs) suggests this is not a rare occurrence. 

\subsubsection{What beliefs do these researchers have about Code LLM\add{s} \del{errors?}}

To better understand how users decide on appropriate uses of Code LLMs in their scientific work, we asked our participants how they think their tool of choice works. Several participants explicitly compared ChatGPT to Google search (P3, P5, P10). P5 said "We're both just racing each other to Google stuff." P10 described ChatGPT "uses a ranking algorithm just like Google- but it summarizes it nicely."  User logs from P14 show the researcher asking ChatGPT to use information on a URL to answer a question about a GitHub repository (at a time when web access was not enabled for ChatGPT). Some interviewees described how information is stored by Code LLMs: P2 referenced ChatGPT's "database". P6 and P9 stated that ChatGPT "memorized" things online. "It's able to use what it has memorized to spit out an answer" (\add{P6}). P4 thought the model might be getting worse by learning from his code.

Two participants described ChatGPT doing calculations or operations: P5 said it performed calculations like Wolfram Alpha. P1 described not needing to run generated code on a personal computer "because it can run Python or R on its own". 

Through logs and descriptions of use cases, there are some indications of attempts to accomplish objectives that typically involve computation or calculation---for example, P6 asks "this [script] is taking too long, what can I do to improve things?" We also observed users asking which machine learning model would run fastest (P10) and asking how many rows would be present in a dataframe after filtering by some condition (P14). It's not clear if these cases represent beliefs about how ChatGPT works, or a willingness to experiment with ChatGPT answers before investing the effort to work on these things. 

We also asked participants what kind of errors they believed to be typical when working with Code LLMs.  "It makes mistakes all the time," says P11. Other participants viewed errors as a rare occurrence: "Occasionally it will make up a function. I'll have to say it doesn't happen often...it's usually handling the data structures and things okay. I think it probably has a good understanding of how to handle different data structures and relate them to each other." (P12) Several participants noted differences in error rates by library/language (P2, P4, P6, P13).

Some participants noted that the correctness of responses depended on the length and complexity of their prompt (P4, P7, P8, P13, P14). However, other participants did not appear to use strategies for simplifying or decomposing multi-step problems before prompting (P1, P10, P11, P12, P14). P11 explained, "...it is just so much more efficient in just mapping out all the steps that you need to do in one go, and then you can just work it out, and otherwise I would have to do it all by myself, think of every single step."  Correspondingly, some user logs showed cases of prompting Code LLMs with hundreds (up to >1000 at the extreme) lines of code.

\section{Discussion}
We now consider the implications of our findings for scientific code practices generally, as well as for designing tools that better support scientific programmers. Throughout this discussion, open-ended directions for further research are identified, and summarized in Table \ref{tab:research-directions}, following a design from \citet{TankelevitchTheAI}'s research roadmap.

\begin{table*}
    \small
    \centering
    \begin{tabular}{p{0.15\linewidth}  p{0.4\linewidth}  p{0.4\linewidth}}
    \toprule
        \textbf{Area} & \textbf{Research questions} & \textbf{Research directions} \\
    \midrule
         Verification strategies & How do scientific programmers determine that generated code is acceptable? & Observational user studies with think-aloud prompts. Eye-tracking to corroborate which code and data elements are used during verification.\\[22pt]
         & Do scientists verify code differently if it's from a Code LLM or a colleague? & Vignette studies asking participants to detect if a code error is present, varying the framing of the code origin.\\[22pt]
         Mental models & How common are incorrect beliefs about the affordances of Code LLMs among scientific users? & Surveys using a scale or open-ended questions about how Code LLMs produce responses. \\[22pt]
         & Do incorrect beliefs about how Code LLMs work predict user behaviors? & Correlate surveyed beliefs about Code LLMs with observed or self-reported behaviors (e.g. number of lines of code accepted per prompt, on average). \\[22pt]
         & Do scientific users correctly perceive the reliability of Code LLMs on a range of relevant tasks? & Assess user predictions of Code LLM reliability on a battery of benchmark tasks and compare user predictions to actual performance. Systematically vary length and complexity of prompts to gauge user's sensitivity to measured relationships with Code LLM performance. \\[22pt]
         & Do scientific users correctly perceive their own ability to detect code errors? & Behavioral experiments to measure actual and perceived rate of detecting code errors on a benchmark set of examples.\\[22pt]
         Programming support & How widespread is adoption of Code LLMs among scientific programmers? & Multi-university surveys, with measures of Chat vs. Copilot interfaces, programming language choice, and demographic information. \\[22pt]
         & How do scientists decide which code tools to use? & Surveys and qualitative interviews about adoption influences. \\[22pt]
         & What programming support are scientists missing, that might be enabled by recent advances in natural language processing? & Design probes for tools that directly aid scientific code translation, documentation generation and retrieval, automatic test generation, and smart highlighting of sensitive parameters. \\
    \bottomrule
    \end{tabular}
    \caption{Future research questions and their candidate methods to better understand the uses, risks, and opportunities for Code LLMs in scientific programming.}
    \label{tab:research-directions}
\end{table*}

\subsection{Supporting scientists who code}

Most scientists we encountered were multilingual, but with mixed fluency in the languages they work in on a daily basis \add{(this has also been noted by \cite{OBrien2024InResearch,Prabhu2011AScience})}. Interviewees describe using Code LLMs as a replacement for documentation and other references, which they find difficult to use in general. This points to a need for more clearly navigable documentation for non-expert programmers or better integration of information retrieval sources into their development environments. Scientific training may not currently provide sufficient education in how to read and use open source software documentation. Furthermore, our interviews raise the possibility that some number of scientific Code LLM users do not realize Code LLMs are not accessing official documentation like a search engine would. Design choices should be explored that enable users to make informed decisions about when to access retrieved documentation versus generations (such as work by \citet{Xu2022In-IDEChallenges}).

\add{If Code LLMs are being used as an aid for scientists working in unfamiliar languages and libraries (primarily through generating or modifying code and answering code-related questions in Chat, in our findings), it is worth considering how these support needs could be met more directly via tools in their developer environment. For example, scientists working with code left behind by a previous lab member could benefit from better documentation of that software, following practices like writing documentation strings for functions and directory README files. Creating this documentation could itself be aided by a Code LLM, but with the supervision of the original author \cite{Ahmed2024CanArtifacts,Dvivedi2024AGeneration}. Such tools may be of particular urgency considering that a recent survey of scientific coders found that less than 30\% reported feeling well-supported to document their programs \cite{Carver2022AStates} (see also \citet{Wiese2020NamingSoftware}). Similarly, scientists who need to translate code between languages may benefit from Code LLMS with interfaces developed explicitly for this purpose \cite{Weisz2021PerfectionTranslation,Weisz2022BetterTranslation}}.

We also noted during exploratory analysis that some programming languages were more associated with Code LLM usage frequency than others. Being a Python programmer had a positive association with usage frequency, whereas being an R, Stata, or SAS user was negatively associated. Being a Python programmer was also correlated with programming in several other languages, suggesting Python programmers may be especially likely to be multilingual (and possibly in need of more support when programming in less familiar contexts). If there are indeed marked differences in Code LLM use by programming language, there are several possible causes. It may reflect the social influence of programming language or scientific communities, with different communities gravitating towards different Code LLM norms. Another possibility is that some languages place lower information retrieval demands on the user---if their syntax and conventions are highly consistent across libraries, or they tend to produce relatively readable error messages. It may be worth considering how Code LLM-support-seeking differs by some measure of within-language consistency or complexity. If there is such a relationship, then that suggests a need for more human-readable programming language interfaces. 

Our observation that Chat users outnumbered Copilot users among survey respondents may be important for designing tools that are compatible with scientist's workflows. On the other hand, this pattern may present distinct disadvantages for programmers: general purpose Chat interfaces can lead users to accept longer, more complex blocks of code \cite{Weber2024SignificantModels}, which means more chances for errors in generations \cite{Kabir2024IsQuestions} and \add{greater cognitive load for the user to verify \cite{Prather2023ItsProgrammers,Barke2023GroundedModels}. Chat may also encourage users to provide lengthy and overly-complex prompts, which is associated with more problematic code generations \cite{Wang2024WhereCode}.}  Additionally, suggestions from Chat must be manually copied to the IDE and translated to work with the scientists' own data structures and variable names (although some interviewees reported they like this step, as it forces them to read the generated code). \add{It would be worthwhile to understand what motivates this choice and whether it is in practice associated with resulting code quality.}

\add{There are several possible explanations for the overwhelming preference for Chat over Copilot in our sample. First, the university where the study was conducted made a large and highly publicized investment in providing a Chat tool to all students, faculty and staff. This study did not directly address the influence of messaging about generative AI, but community attitudes are known to factor into programmers' beliefs about using the usability and reliability of Code LLMs \cite{Russo2023NavigatingEngineering,Cheng2024ItTools,Padiyath2024InsightsCourse}. Additionally, scientists may not seek out developer tools \cite{Wilson2016SoftwareLearned}, so perhaps many are unaware of in-IDE Copilots. Another possibility is privacy concerns: our sample includes many biomedical researchers, so a university-endorsed Chat tool may be judged safer for programming that interacts with potentially sensitive data. Surveys (ideally at multiple universities and research institutes) should directly assess how scientists learned about their Code LLM of choice, alternatives considered, and how they decided between them.}

\subsection{Implications for scientific code verification}
Most participants responded to questions \add{about their experiences with problematic code generations} with confidence that they could verify Code LLM outputs, though there were exceptions (like P8 and P13). Even though we did not design our study to seek out verification failures, we spotted several cases from user logs where successful verification did not occur. Interestingly, all three such cases occurred when researchers were using ChatGPT rather than Copilot, and involved prompting the model to interpret, generate, or modify relatively long blocks of code (40 lines or more). These were also cases where the user prompts did not decompose problems, but contained compound instructions (i.e., "generate a signal processing pipeline with several steps", "add a feature and return the updated code") or ambiguity ("which code should I use?"). 

\add{The verification strategies surfaced here were consistent with strategies reported in other studies: inspecting data artifacts produced by code, reading programs line by line, and pattern-matching have by now been observed in several user groups \cite{Barke2023GroundedModels,Prather2023ItsProgrammers,McNutt2023OnNotebooks,Ragavan2022GridBook:Grid,Vaithilingam2022ExpectationModels,Liang2024AChallenges}. Some participants also described using generated code explanations as reference material to decide if the code is correct. This is congruent with studies of beginner programmers, who are prone to overconfidence in both generated code and code explanations \cite{Prather2023ItsProgrammers,Nguyen2024HowOther,Liu2023WhatModels} (as well as other forms of automatic feedback, like autograders \cite{Lee2011PersonifyingLearning, Keuning2018AExercises})}.

\add{In the context of scientific programming, where code must correspond to valid experimental and analytical decisions, the verification strategies reported by participants are imperfect. "Eyeballing" data artifacts created by a program scales poorly as data  grows in number of features, observations, and complexity of possible values. Similarly, reading code line-by-line becomes onerous as more lines are produced \cite{Prather2023ItsProgrammers, Barke2023GroundedModels}. Other strategies, like referring to the generated code explanations, are vulnerable because they are circular. Until a user checks an explanation against some external documentation, it is unverified--but many interviewees  used Chat precisely because they could \textit{not} use documentation.}

While our study design does not allow us to estimate what percent of errors in generated code researchers might overlook, we can consider our findings in light of a study by \citet{Kabir2024IsQuestions}, in which programmers reviewed ChatGPT-generated answers to Stack Overflow questions. In about half of ChatGPT's answers, incorrect information was present, and a sample of users with moderate programming experience failed to identify incorrect aspects in 39\% of answers. While it is hard to directly compare the programming experience of our sample to theirs, our participants who use ChatGPT (or similar browser-based chat interfaces) possibly ask similar questions as those in the Stack Overflow benchmark set (indeed, several interviewees noted  that ChatGPT had replaced Stack Overflow, Google search, and library documentation). If a 39\% miss rate is roughly expected, then the common perception among our interviewees who use ChatGPT--that they typically could catch errors in code generations and accompanying explanations---could be miscalibrated.

\subsection{The role of mental models}
Like other investigations of non-expert programmers using Code LLMs \cite{Prather2023ItsProgrammers,Nguyen2024HowOther,Liu2023WhatModels}, we observe that some users have incorrect mental models of how LLMs generate responses. In particular, several participants described ChatGPT as a search engine, and some participants either used it for calculations or stated that it was capable of performing computations similar to a calculator. If a user believes that ChatGPT accesses the documentation for a software library similar to a Google search, then it seems reasonable that users would consider it a complete substitute for documentation. It would be interesting to study to what extent misunderstandings of how Code LLMs work relates to verification practices, both among scientists and the general population of non-expert users.

\add{It may be important to disambiguate overconfidence in the Code LLM from overconfidence in a user's ability to detect errors (for example, as conceptualized by \citet{TankelevitchTheAI}). If users overestimate the reliability of Code LLM solutions for a given task, this may be improved via interface cues (e.g. confidence highlighting \cite{Weisz2021PerfectionTranslation}) or behavioral guardrails (e.g. detecting when a user is asking a question that involves algebraic operations, and alerting the user that Code LLMs do not work like calculators). Alternatively, users who fully understand that Code LLMs are error-prone but misunderstand their own ability to spot errors may need reminders to limit the size of code blocks accepted at once, or tools that give them better visibility into what code does \cite{Nam2024UsingUnderstanding, Liu2023WhatModels}.}

\add{To determine whether users are well calibrated to the actual reliability of Code LLMs, researchers might ask users to estimate what percent of the time the Code LLM will produce an acceptable result for various prompts from an existing benchmark study (such as \cite{Wang2024WhereCode, Kabir2024IsQuestions}). Experimenters should vary prompt dimensions like length and task complexity to probe if users are sensitive to these established effects. The difference between the user's predictions of Code LLM reliability and actual would give a measure of how well calibrated the user is to the model's error patterns. Such a measure might also reveal if users treat Code LLMs as uniformly reliable across tasks, or if they perceive a more jagged landscape of competencies (which would be expected for more experienced users \cite{Sarkar2022WhatIntelligence,Liu2023WhatModels})} 

\add{To study how overconfident users are in their verification abilities, behavioral studies asking programmers to determine if an error is present in a code block could be conducted with variable code lengths and complexities. Participants should rate their confidence under different conditions to assess if they are appropriately calibrated to their actual performance. Vignettes could be introduced to test if programmers apply the same scrutiny whether code is said to originate from a colleague or a Code LLM.}

\subsection{Scientific software and knowledge}

Because we lack a clear baseline for how common code errors are in science research, and what forms they commonly take, it is hard to know if adoption of Code LLMs will substantially change the rate or severity of serious errors---and consequently, the quality and trustworthiness of new scientific knowledge. Analyzing retraction notices associated with code errors may provide clues into the nature of typical errors historically, but this sampling technique would only account for a small fraction of cases. Scientific code errors may not be easy to identify by, for example, testing that scientific code runs or produces a specific output, because in many cases the correct output is not knowable in advance. 

It is important to consider that not all code errors are equally problematic. For example, in the case of a scientist using ChatGPT to generate code for signal processing and receiving code with the wrong window size, we can't conclude this would lead the researcher to a radically different interpretation of their data. In fact, it is often desirable for scientific results to be should be robust to many analysis choices---this is a premise of "multiverse analysis" \cite{Sarma2023Multiverse:Notebooks,Gu2024HowAnalyses, Dragicevic2019IncreasingAnalyses,Steegen2016IncreasingAnalysis}, in which experimenters deliberately explore the robustness of results to a range of reasonably-chosen parameters in analysis code. A challenge for scientific programmers is that they may not realize when a Code LLM suggestion modifies parameters beyond their reasonable range. Similarly over-reliance on generated code increases the risk of making unintentional experimental design choices ("How do we handle participants who fail an attention check?") when a user believes they are making only programmatic choices. \add{One research direction for scientific code support could be tools that automatically identify code blocks and parameters whose variance has critical ramifications for results. This could be supported via domain-specific knowledge bases and numerical simulations (e.g., simulating the effect of varying a parameter value on the output of a program, a kind of sensitivity analysis \cite{Oakley2004ProbabilisticApproach,McCartan2022Adjustr:Sampling}).}

A challenge for reproducibility in future scientific work may be that if scientists can't entirely verify the behavior generated code, published writing \textit{about} their code in scientific manuscripts may not closely match the code they actually have. It will become especially important, then, for code to serve as its own documentation---in other words, expanding the code-sharing norms that have already begun in many scientific fields \cite{Nosek2024ReplicabilityScience, NationalAcademiesofSciencesandPolicyandGlobalAffairsandBoardonResearchDataandInformationandDivisiononEngineeringandPhysicalSciencesandCommitteeonAppliedandTheoreticalStatisticsandBoardonMathematicalSciences2019ReproducibilityScience,Munafo2017AScience}.

\add{Existing communities that are professionalizing research software and up-skilling scientific developers \cite{Wilson2012BestComputing,Wilson2016SoftwareLearned,Sims2021ResearchIdentity,Carver2013Self-perceptionsEngineers} may be an ideal locus for discussions about what Code LLM practices should be acceptable in science, what code quality-checking mechanisms should exist, and how to advocate for informed practices and standards. In particular, this may be a critical time to revisit conversations about code testing, a practice that has long been supported by scientific coders \cite{Carver2022AStates, Prabhu2011AScience} but remains irregularly implemented. There may also be roles for Code LLMs in test development: AI-supported automatic test-development is an active area of research \cite{Schafer2024AnGeneration,Ebert2023GenerativePractitioners,Bhatia2024UnitTools}, although these efforts would likely have to be adapted to fit scientific contexts.} 

\subsection{Study Limitations} 
Several cautions must be made when interpreting our results. First, both our survey and interviews rely on self-reports of behavior, which may not match actual interactions with Code LLMs \add{(for example, users may report completing coding tasks faster with using Code LLMs despite objective measures otherwise \cite{Vaithilingam2022ExpectationModels,Prather2024TheProgrammers})}. Aside from reviewing a sample of user logs made available by the participants, our study does not allow us to directly observe participants programming with or without Code LLMs. Second, there are inherent limitations from our sampling strategy. We use a convenience sample of researchers at one well-resourced university in the United States. These individuals may have access to more training, both in programming and Code LLM usage, than is typical in the scientific community. Third, due to the time it takes to participate in a study, our sample likely skews towards individuals with more flexible schedules (potentially earlier-career scientists) or who are especially interested in programming. Finally, our survey findings in particular should be considered exploratory, rather than confirmatory: major findings should be replicated in a larger sample with better coverage of other research environments in order to make strong inferences about scientific practice generally.

\section{Conclusion}
In total, our results indicate that many scientists who program with Code LLMs use general purpose, browser-based chat interfaces like ChatGPT, rather than in-IDE assistants like Copilot. Many of these programmers come from backgrounds outside historically computational fields like Computer Science or Engineering. Through interviews, we learn that scientists often use Code LLMs as a kind of documentation for unfamiliar programming languages and libraries they need to learn on the job. Verification strategies typically reported involve running code and inspecting the output visually, reading code suggestions line by line, looking for high-level patterns in code, and referencing Code LLM-generated explanations for suggestions. We observed cases were these strategies were not successful and Code LLMs introduced unplanned or incorrect elements into research code. We also encountered a variety of misconceptions about how Code LLMs work. These findings suggest that to avoid degrading scientific code, Code LLM users may need clearer signals about the kinds of errors Code LLMs can make, as well as interfaces that allow them to intentionally choose between information retrieval and generation. 

\begin{acks}
This work was made possible by a grant from the Alfred P. Sloan Foundation.
\end{acks}

\bibliographystyle{ACM-Reference-Format}
\bibliography{references}

\end{document}